# PEG- and PVP-assisted wet-chemical synthesis of ZnS nanocrystals via hydrothermal and co-precipitation methods for route-dependent structural and bandgap tuning

R. U. AHMAD[a,*], N. JAVED[a], F. SHER[b]

[a] *Faculty of Basic Sciences, Ghulam Ishaq Khan Institute of Engineering Sciences and Technology, Topi 23640, Pakistan*

[b] *Department of Materials Science and Engineering, University of Illinois Urbana-Champaign, USA*

*Corresponding author: raouzair854@gmail.com; e-mails: nasir.javed@giki.edu.pk; falaks2@illinois.edu

Zinc sulphide (ZnS) nanocrystals were prepared by room-temperature co-precipitation and hydrothermal synthesis using polyvinylpyrrolidone (PVP) and polyethylene glycol (PEG) as growth-controlling agents. X-ray diffraction confirmed cubic zinc-blende ZnS. Polymer addition reduced the XRD-derived crystallite size, reaching 2.03 nm for 0.50 mM PVP, compared with 3.33 nm for PEG. Hydrothermal samples showed larger crystallites of 5.41–14.31 nm. The optical band gap of co-precipitated ZnS increased from 3.62 to 3.81 eV with decreasing crystallite size, whereas hydrothermal samples remained near 3.61 eV. FTIR, SEM and DLS results supported polymer association and reduced aggregation.

## 1. Introduction

Nanomaterials have revolutionized modern science, offering unique properties such as a larger surface area, enhanced reactivity, and size-dependent behaviour that are not found in bulk materials [1]. These properties arise due to their small size, typically in the range of a few nanometers, which results in quantum mechanical effects dominating their behaviour. Among these nanomaterials, quantum dots (QDs) stand out as a particularly promising class.

The QDs have numerous advantages over bulk materials that lie primarily in their ability to tune the bandgap by controlling their size. In bulk semiconductors, the energy bands are continuous, limiting their ability to absorb or emit specific wavelengths. On the other hand, QDs, with their discrete energy levels, can be engineered to absorb and emit light across the visible and near-infrared spectrum, depending on their size [2]. This property makes QDs highly desirable for applications such as high-efficiency solar cells, where capturing a wider range of the solar spectrum is critical [3]. Furthermore, the quantum confinement effect leads to enhanced optical properties, such as high photoluminescence efficiency and strong absorption coefficients, which bulk materials cannot match [4]. Additionally, QDs can be incorporated into flexible and lightweight devices, offering significant advantages in terms of miniaturization and versatility compared to traditional materials [5].

Most of the research and commercial applications in the field of quantum dots are dominated by Cd and Pb-based materials and their derivatives[6–8]. Cadmium-based quantum dots, such as CdSe and CdTe, have been extensively studied and utilised due to their exceptional optical properties, tunable bandgaps, and easy synthesis, which make them suitable for a wide range of applications, including optoelectronics and photovoltaics [9,10]. However, the widespread use of these materials raises significant concerns regarding their environmental and health impacts. Cadmium-based QDs, while efficient, pose significant ecological risks due to bioaccumulation, with studies showing even trace amounts ($\leq$ 0.1 ppm) can disrupt aquatic ecosystems [11]. Efforts are underway to develop alternative, environmentally benign QD materials with lower toxicity to mitigate these concerns while maintaining or improving performance characteristics [12]. Among these alternatives, ZnS has emerged as a leading material due to its wide bandgap (~3.6 eV), low toxicity, and high thermal stability [13,14]. ZnS QDs are particularly useful in environmentally sensitive applications such as solar cells [15], light-emitting diodes (LEDs) [16], and bioimaging [17], where both optical performance and biocompatibility are essential. The potential of ZnS QDs in these applications is immense, particularly for energy harvesting through photovoltaics, where they can enhance the efficiency of solar cells by enabling UV absorption [18]. In addition, their photocatalytic properties make ZnS QDs effective for environmental remediation, such as degrading organic pollutants under UV light [19]. Biocompatible ZnS QDs also hold promise as safe contrast agents for imaging applications [20], including magnetic resonance imaging (MRI) [20] and fluorescence-guided surgery [21]. However, despite their promising properties, the synthesis of high-quality QDs remains a challenge. Achieving uniform particle size, high crystallinity, and minimal defects is essential for maximizing the optical and electronic performance of ZnS QDs [22–24]. Several synthesis methods have been developed to produce QDs, including traditional plasma synthesis, epitaxial growth techniques, hot injection methods, hydrothermal methods, and co-precipitation techniques [25–27]. Each of these methods has its advantages, but they also come with significant challenges.

Plasma synthesis methods, for instance, often yield high-quality QDs with precise size control but typically require high-power arcs and struggle with purity[28]. On the other hand, epitaxial growth methods deposit high-quality, precisely controlled QDs directly onto the substrate, which is ideal for device integration. Still, it is very energy-intensive and time-consuming [29]. The hot injection method is a solution processing technique that provides high purity and precise size control. Still, it uses toxic solvents such as trioctylphosphine oxide (TOPO), which is not environmentally friendly[30][31]. This range of synthesis methods presents a variety of trade-offs in terms of purity, environmental impact, and process complexity, highlighting the ongoing challenges in producing ZnS QDs that meet stringent performance and sustainability criteria.

Hydrothermal method, on the other hand, is considered eco-friendlier and more scalable because it can be performed at moderate temperatures and pressures in aqueous solutions. However, controlling the size and uniformity of the QDs synthesised using hydrothermal method can be challenging, and the process often results in larger particle sizes and aggregates, which can negatively impact the performance of the resulting QDs. Sabaghi et al. reported that the hydrothermal synthesis of ZnS nanoparticles produces particles of size up to 50 nm [32]. While the hydrothermal method has the potential to yield high-quality QDs, challenges remain in controlling particle size and morphology. The reaction time, precursor concentration, and temperature need to be carefully optimised to obtain uniform particles [33].

On the other hand, the co-precipitation method operates at room temperature and is considered environmentally friendly; it typically yields a wide distribution of particle sizes [34]. In a co-precipitation synthesis, the metal precursor is dissolved in deionized water. This produces high-purity nanoparticles with very little energy consumption. But this method struggles with a wider size distribution. These issues highlight the need for further optimisation of synthesis conditions to achieve ZnS QDs with the desired properties for various applications [34]. In co-precipitation synthesis, size control is achieved by adjusting parameters such as reaction time and ligand-to-precursor ratios. For example, extended reaction times typically result in larger particles due to Ostwald ripening, a process in which smaller crystals dissolve and redeposit onto bigger ones [35][36].

In this study, we aim to address these challenges by optimizing the hydrothermal and co-precipitation synthesis method for ZnS QDs. Our goal is to fine-tune the reaction parameters, such as precursor ratios and capping agent concentrations, to achieve ZnS QDs with uniform size distributions and enhanced optical properties. By overcoming the limitations of the current synthesis methods, this work aims to contribute to the development of high-quality ZnS QDs that can be integrated into practical devices, advancing their use in fields such as photovoltaics, optoelectronics, and biomedical applications. Ultimately, the findings from this study will provide valuable insights into the synthesis and application of ZnS QDs, offering a pathway to scalable and sustainable synthesis methods.

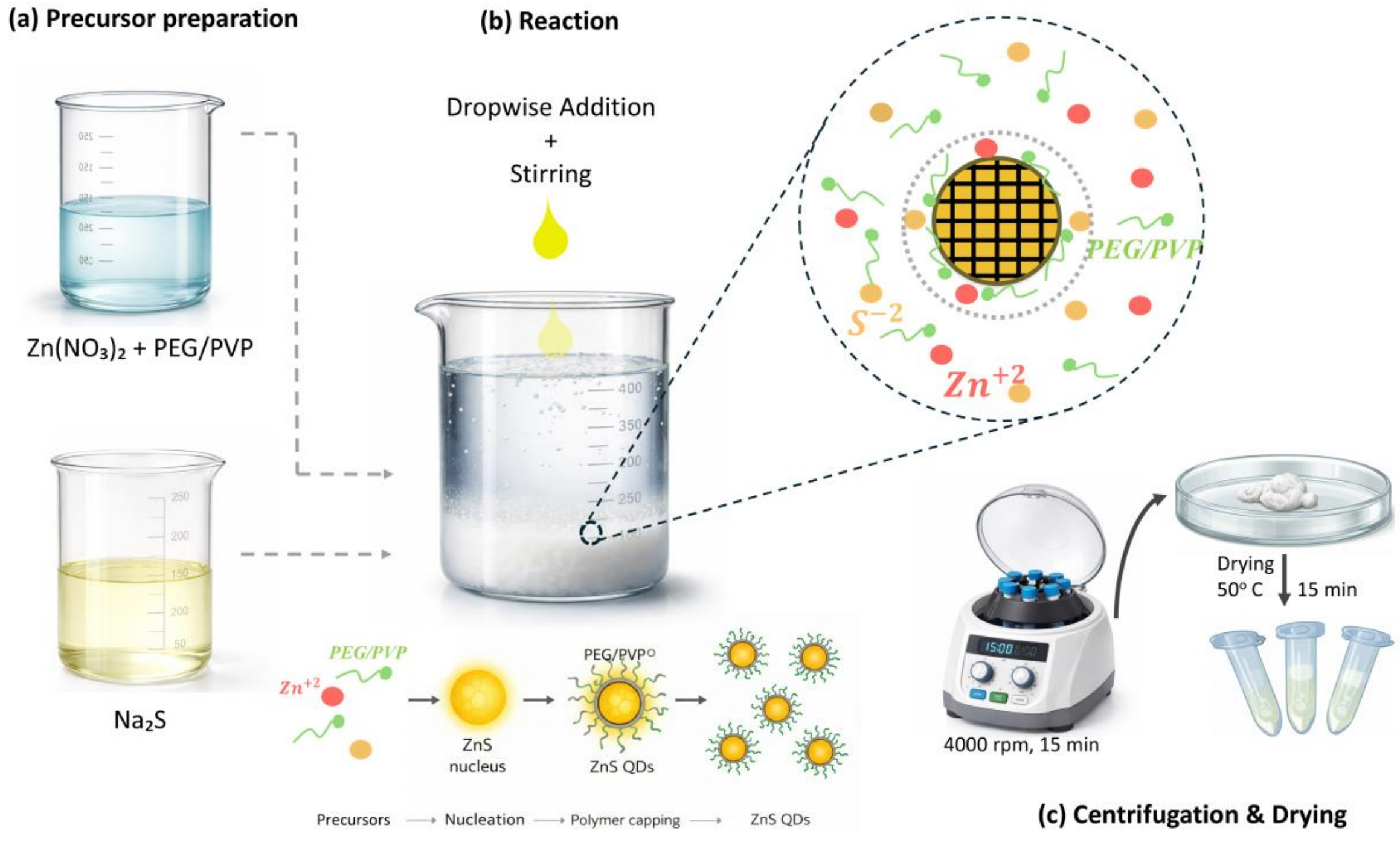

*Fig. 1. Schematic illustration of the co-precipitation synthesis route for ZnS QDs using PVP and PEG as capping agents. The stepwise procedure includes precursor dissolution, controlled dropwise mixing, and post-synthesis washing and drying*

## 2. Experimental

### 2.1. Materials

All materials used in this study were of laboratory grade. Zinc nitrate hexahydrate ($Zn(NO_3)_2 \cdot 6H_2O$ purity $\geq$ 99.0%) and polyvinylpyrrolidone (PVP, average molecular weight = 40,000 g/mol) were purchased from Sigma-Aldrich. Sodium sulphide trihydrate ($Na_2S \cdot 3H_2O$) was obtained from VWR Chemicals, while polyethylene glycol (PEG average molecular weight = 6,000 g/mol) was acquired from Fluka. All materials were used as acquired without any further purification.

### 2.2. Methods

Co-precipitation Synthesis: The Synthesis method is schematically depicted in Fig. 1. Briefly, 0.9520 g of zinc nitrate hexahydrate ($Zn(NO_3)_2 . 6H_2O$) were dissolved in 80 ml of deionized (DI) water under constant stirring. The solution was then added with 0.024 g (0.05 mM) of PEG. This solution was continuously stirred for 10 minutes at 500 rpm using a magnetic stirrer, resulting in a transparent solution. A similar solution was prepared by dissolving 0.4181 g of sodium sulphide in 80 ml of DI water. This solution was also stirred continuously at 500 rpm for 10 minutes, yielding a very light-yellow solution. The prepared sodium sulphide solution was transferred into a burette, from where it was added to the zinc nitrate + PEG solution dropwise at a speed of around 150 drops per minute. The solution was added under continuous stirring to ensure uniform agitation. This turns the transparent solution into a turbid white dispersion with precipitates. This dropwise addition of $Na_2S$ prevents uncontrolled nanoparticle growth and leads to uniform particle size. This mixture was then left for half an hour at room temperature to give it time to react. The chemical reaction is described in the following chemical equation (1).

$$Zn(NO_3)_2 . 6H_2O + Na_2S \xrightarrow{aqueous} ZnS + 2NaNO_3 \quad (1)$$

The turbid white dispersion was centrifuged at 4000 rpm for 15 minutes, and the precipitates were collected. The precipitates were then washed several times with DI water and ethanol (4000 rpm for 15 minutes each cycle) to remove any unreacted precursors. The resulting precipitate was collected and dried at 50°C for 12 hours in an oven. After drying, the final product was ground into a fine powder and stored for further characterisation. The same procedure was followed for the synthesis using 0.048 g (0.10 mM) and 0.24 g (0.50 mM) of PEG. For the second batch with PVP as the capping agent, the same molar concentrations (0.05 mM, 0.10 mM, 0.50 mM) were used, with masses of 0.16 g, 0.32 g, and 1.6 g.

Hydrothermal Synthesis: The synthesis starts with the preparation of precursor solutions, as shown in **Fig. 2**, where zinc sulphate pentahydrate and sodium sulphide are dissolved separately in deionized water and stirred thoroughly to form $Zn^{2+}$ and $S^{2-}$ ions. The sodium sulphide solution is then slowly added to the zinc solution under constant stirring, allowing ZnS nanoparticles to form. The resulting mixture is transferred into a Teflon-lined stainless-steel autoclave, filled to half its capacity, and heated at 220 °C for 12 hours to promote uniform crystallization under high pressure and temperature. After natural cooling to room temperature, the product is centrifuged, and the precipitate is washed several times with water and ethanol to remove impurities as described above. The purified ZnS nanoparticles are then dried at 50 °C for 12 hours, yielding a fine powder ready for further characterisation.

### 2.3. Characterisation

Scanning electron microscopy (SEM) analysis was performed using a ZEISS EVO 15 instrument integrated with energy-dispersive X-ray spectroscopy (EDX) to examine the morphology and elemental composition of the samples. X-ray diffraction (XRD) patterns were recorded using an AXRD-LPD Powder X-ray Diffractometer with Cu Kα radiation to study the structural properties and the crystallite size estimation. The absorption spectra were recorded with the Shimadzu UV-1800 double-beam spectrophotometer. An IRTracer-100 Fourier Transform Infrared (FTIR) Spectrometer equipped with InGaAs, DLATGS, and MCT detectors was used to confirm the surface functionalisation of the sample. Dynamic light scattering measurements (DLS) were used to calculate the hydrodynamic radius of the particle and the particle size distribution. Malvern Panalytical's Zetasizer Blue was used for this purpose.

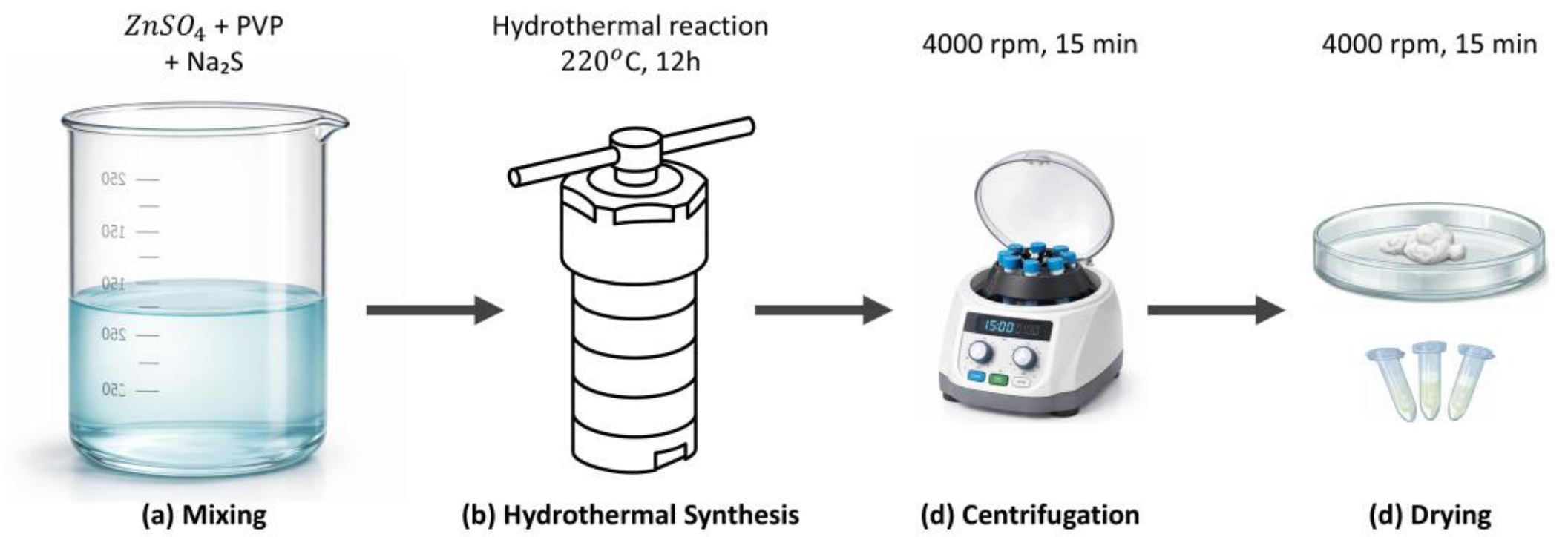

*Fig. 2. Schematic of the hydrothermal synthesis route for ZnS QDs, showing precursor preparation, autoclave reaction, and purification*

## 3. Results and discussion

### 3.1. Structural characterisation and crystallite size

The structure of the synthesised ZnS QDs was characterised by powder X-ray diffraction (XRD). Fig. 3 shows representative XRD patterns for ZnS QDs synthesised by the co-precipitation method. All observed peaks match the standard zinc-blende ZnS reference (PDF #05-0566), confirming the formation of pure cubic ZnS. For the PEG-capped QDs (Fig. 3a), strong Bragg reflections appear at 2θ ≈ 28°, 47°, 56°, and 76°, corresponding to the (111), (220), (311), and (400) lattice planes, respectively. By applying Bragg's law, we calculated the interplanar spacing *d* and lattice constant *a* for each sample (). The lattice constants in all cases are very close to the bulk ZnS value (*a* ≈5.41 Å), with only slight compressive shifts reflecting nanoscale effects.

Introducing PVP as the capping agent produces essentially the same set of diffraction peaks (Fig. 3b), but with slightly different parameters. In fact, at each polymer concentration, the PVP-capped samples exhibit marginally smaller d-values and lattice constants than the analogous PEG-capped ones (). For example, at 0.50 mM polymer, the (111) *d*-spacing is ~3.10 Å for PVP versus ~3.12 Å for PEG, giving *a* ≈5.36 Å (PVP) vs ~5.41 Å (PEG). As the polymer concentration increases, the interplanar spacing decreases, leading to a reduction in the lattice constant from the standard value. This indicates that higher capping agent concentrations result in greater compression of the particles. Also, the large PVP molecules bind strongly to the ZnS surface and arrest crystal growth more effectively than PEG.

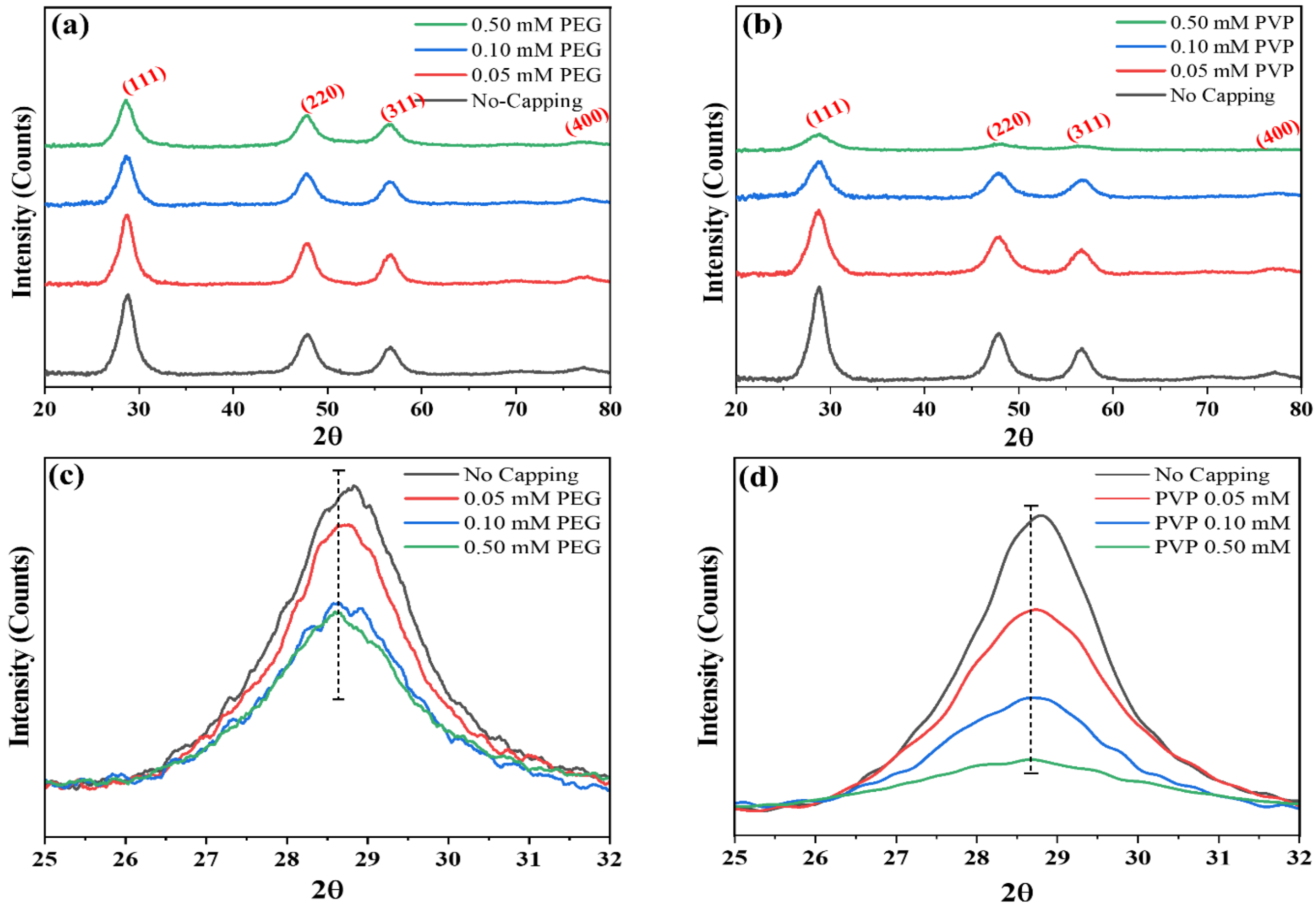

*Fig. 3. XRD patterns of ZnS QDs synthesised by the co-precipitation method with (a) varying PEG concentrations, (b) varying PVP concentrations. (c) and (d) are corresponding enlarged (111) peaks*

*Table 1. Interplanar spacings and lattice parameters of ZnS nanocrystals prepared by co-precipitation with different concentrations of PVP and PEG.*

| **Sample** | **Plane (hkl)** | **Planer Spacing ($d_{hkl}$) (Å) Standard $d_{111}$ = 3.12** | **Lattice constant ($a_{hkl}$) (Å)** | **$\bar{a}$ (Å) Standard $d$ = 5.41** |
|---|---|---|---|---|
| Co-Precipitation Synthesis | | | | |
| 0.05 mM PVP | (111) | 3.10 | 5.37 | 5.37 |
| | (220) | 1.89 | 5.36 | |
| | (311) | 1.62 | 5.38 | |
| 0.10 mM PVP | (111) | 3.10 | 5.38 | 5.37 |
| | (220) | 1.90 | 5.37 | |
| | (311) | 1.62 | 5.37 | |
| 0.50 mM PVP | (111) | 3.10 | 5.38 | 5.36 |
| | (220) | 1.88 | 5.32 | |
| | (311) | 1.63 | 5.41 | |
| 0.05 mM PEG | (111) | 3.14 | 5.45 | 5.44 |
| | (220) | 1.92 | 5.44 | |
| | (311) | 1.64 | 5.44 | |
| 0.10 mM PEG | (111) | 3.13 | 5.42 | 5.41 |
| | (220) | 1.91 | 5.41 | |
| | (311) | 1.63 | 5.41 | |
| 0.50 mM PEG | (111) | 3.12 | 5.41 | 5.41 |
| | (220) | 1.91 | 5.41 | |
| | (311) | 1.63 | 5.42 | |

A noticeable change in the full width at half maximum (FWHM) of the XRD peaks was observed with varying polymer concentration. This variation in FWHM reflects changes in the crystallite size of the material, as explained by the Scherrer and Williamson-Hall (W-H) methods. W-H plots are used to decouple the contributions of crystallite size and microstrain to XRD broadening [37]. For example, Goswami applied W-H analysis to PVP-capped ZnO, extracting a crystallite size of ~17 nm and a strain of ~1.0 $\times$ $10^{-3}$ (for uncapped) versus ~12 nm and 2.89$\times 10^{-3}$ for capped ZnO [38]. To accurately determine the peak positions (2θ) and FWHM values, Gaussian fitting was applied to each diffraction peak. These parameters were then used to estimate the crystallite size using the Scherrer formula, and the lattice strain was assessed through the W-H analysis. In the W-H approach, a plot of $\beta cos\theta$ versus $4sin\theta$ was constructed for all observed peaks. A linear fit was applied using the Gaussian fitting tool in OriginLab to determine the slope and y-intercept of the line. These values were subsequently used to calculate the crystallite size and the lattice strain [37].

The crystallite size of uncapped ZnS QDs is 4.51 nm. The capping reduced the crystallite size of the QDs significantly to 2.03 nm with the capping of 0.50 mM PVP and to 4.03 nm with 0.05 mM PEG capping. The increase in FWHM of the peaks can be clearly seen in Fig. 3c and Table 2. This follows the same trend as Ajitha et al. found PVP-capped silver nanoparticle crystallites of only 26 nm (Scherrer size) versus 39 nm for PEG-capped [39]. The W-H analysis demonstrated that higher capping agent concentrations introduced more strain into the ZnS lattice, especially for PVP, where strain values reached up to 18 × $10^{-3}$. This result aligns with findings by Wiranwetchayan et al., who report microstrain ε ~ 0.33 $\times$ $10^{-3}$ for $TiO_2$-PVP versus 0.71 $\times$ $10^{-2}$ for $TiO_2$-PEG [40]. Even low PVP concentrations can cause strain. Soltani et al. found ε ~ 4.5 $\times$ $10^{-3}$ for 0.08 mM PVP [41]

*Table 2. XRD peak widths, Scherrer crystallite sizes, Williamson–Hall crystallite sizes and estimated microstrain values for co-precipitated ZnS samples. Crystallite sizes represent coherent diffraction domains and should not be interpreted directly as individual particle diameters.*

| **Sample** | | **PVP** | | | | | **PEG** | | | | |
|---|---|---|---|---|---|---|---|---|---|---|---|
| | | Scherrer Method | | | W-H Method | | Scherrer Method | | | W-H Method | |
| | Planes | FWHM | $D_{hkl}$ | $D_{avg}$ | D | E ($10^{-3}$) | FWHM | $D_{hkl}$ | $D_{avg}$ | D | ε ($10^{-3}$) |
| No Capping | (111) | 2.047 | 4.01 | 4.51 | 4.38 | 0.27 | | | - | - | - |
| | (220) | 1.888 | 4.60 | | | | - | - | | | |
| | (311) | 1.838 | 4.91 | | | | | | | | |
| 0.05 mM | (111) | 2.459 | 3.33 | 3.35 | 3.05 | 2.56 | 2.047 | 4.00 | 4.03 | 3.95 | 0.52 |
| | (220) | 2.862 | 3.04 | | | | 2.149 | 4.04 | | | |
| | (311) | 2.443 | 3.69 | | | | 2.222 | 4.06 | | | |
| 0.10 mM | (111) | 2.596 | 3.16 | 3.24 | 2.94 | 2.91 | 2.132 | 3.84 | 3.89 | 3.60 | 1.87 |
| | (220) | 2.728 | 3.18 | | | | 2.333 | 3.72 | | | |
| | (311) | 2.657 | 3.39 | | | | 2.190 | 4.11 | | | |
| 0.50 mM | (111) | 3.464 | 2.36 | 2.03 | 3.36 | 18.3 | 2.202 | 3.72 | 3.33 | 4.71 | 8.41 |
| | (220) | 4.633 | 1.87 | | | | 2.738 | 3.17 | | | |
| | (311) | 4.802 | 1.87 | | | | 2.882 | 3.12 | | | |

It is observed that the size reduction was more pronounced in the case of PVP capping as compared to PEG, indicating a stronger interaction of PVP with the ZnS surface. This is because of the amphiphilic nature of PVP [42]. In PVP, the hydrophilic pyrrolidone ring functions as the head group, while the hydrophobic polyvinyl chain serves as the tail. PVP plays a dual role in the synthesis process, (i) it can regulate particle growth by forming a passivation layer around the ZnS core through coordination interactions between the nitrogen atom in the pyrrolidone group and $Zn^{+2}$ ions, and (ii) it helps prevent particle agglomeration through steric hindrance, which arises from repulsive forces among the polyvinyl chains. As a result, the encapsulation by PVP provides a confined environment around the ZnS nanocrystals, influencing their growth and stability.

ZnS QDs were also synthesised via the hydrothermal method. These samples exhibited sharper and more intense diffraction peaks (shown in Fig. 4) compared to those from the co-precipitation process, indicating larger crystallite sizes and improved crystallinity. All observed peaks correspond to the (111), (220), and (311) planes of cubic zinc-blende ZnS (PDF# 05-0566), confirming successful phase formation without any secondary phases.

The calculated lattice constants for all three hydrothermal samples were very close to the standard bulk ZnS value of 5.41 Å, as shown in Table 3. This minimal deviation suggests a relaxed crystal structure with negligible internal stress. Unlike the co-precipitation samples, where increasing polymer concentration led to noticeable lattice compression, the hydrothermal method yielded ZnS crystals with bulk-like lattice parameters, reflecting the thermodynamic stability achieved under high-temperature and pressure conditions.

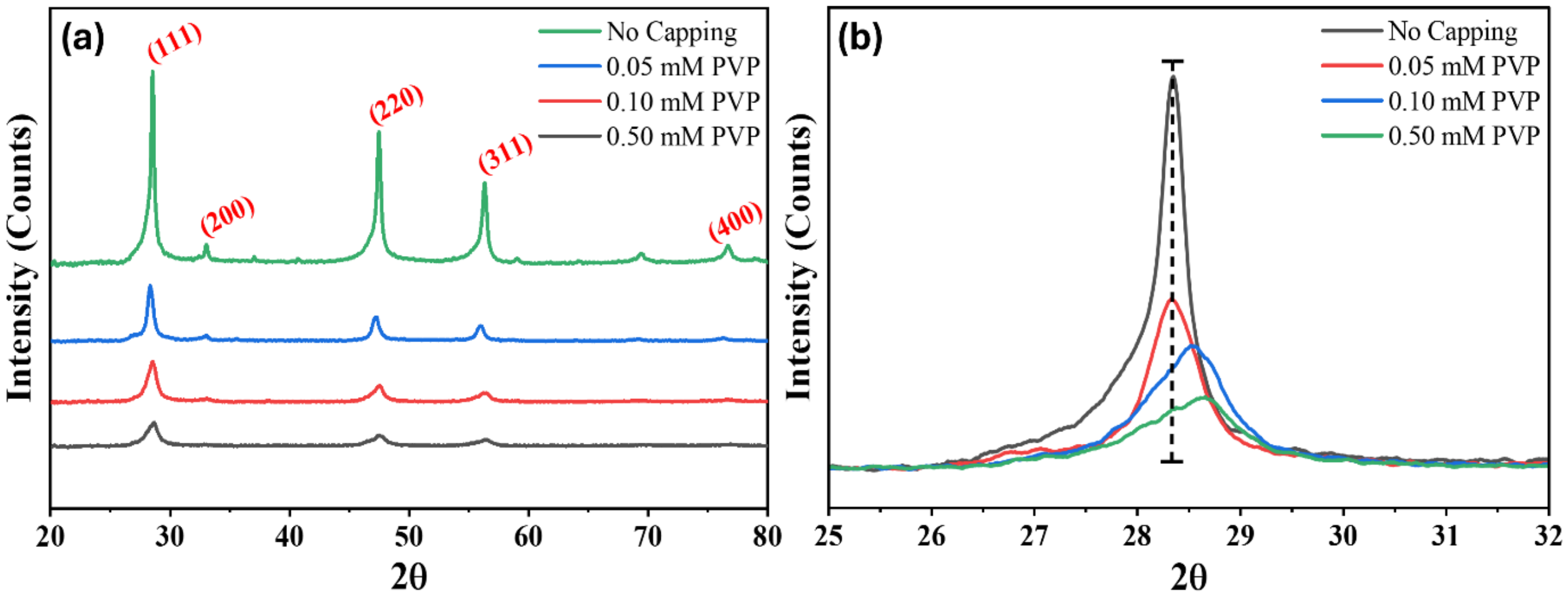

*Fig. 4. (a) XRD spectra of ZnS QDs synthesised by the hydrothermal method and (b) the corresponding enlarged (111) peak.*

*Table 3. Interplanar spacings and lattice parameters of ZnS nanocrystals prepared by hydrothermal synthesis with different PVP concentrations.*

| **Sample** | **Plane (hkl)** | **Planer Spacing($d_{hkl}$) (Å) Standard $d_{111}$ = 3.12** | **Lattice constant($a_{hkl}$) (Å)** | **$\bar{a}$ (Å) Standard = 5.41** |
|---|---|---|---|---|
| Hydrothermal Synthesis | | | | |
| No Capping | (111) | 3.12 | 5.42 | 5.41 |
| | (220) | 1.91 | 5.41 | |
| | (311) | 1.63 | 5.41 | |
| 0.05 mM PVP | (111) | 3.12 | 5.45 | 5.44 |
| | (220) | 1.92 | 5.44 | |
| | (311) | 1.63 | 5.44 | |
| 0.10 mM PVP | (111) | 3.12 | 5.42 | 5.41 |
| | (220) | 1.91 | 5.41 | |
| | (311) | 1.63 | 5.41 | |
| 0.50 mM PVP | (111) | 3.11 | 5.41 | 5.41 |
| | (220) | 1.91 | 5.41 | |
| | (311) | 1.63 | 5.42 | |

To quantify crystallite size and strain, both the Scherrer equation and Williamson–Hall (W–H) analysis were employed. The Scherrer method gave average crystallite sizes in the range of ~5.41 nm to 14.6 nm. W-H plots, which decouple the effects of size and strain on peak broadening, revealed slightly larger diameters along with strain values between $1.02 \times 10^{-3}$ and $4.31 \times 10^{-3}$ (Table 4). These strain values at higher PVP concentrations are significantly lower than those in the co-precipitated samples, supporting the observation of sharper diffraction peaks and improved lattice ordering. Introducing the PVP in the hydrothermal process significantly decreased the crystallite size to 5.41 nm for 0.50 mM PVP (via Scherrer method). The W-H analysis shows an increased lattice strain of $4.31 \times 10^{-3}$, supporting the co-precipitation results.

*Table 4. XRD peak widths, Scherrer crystallite sizes, Williamson–Hall crystallite sizes and estimated microstrain values for hydrothermally synthesised ZnS.*

| **Sample** | **Scherrer Method** | | | | **W-H Method** | |
|---|---|---|---|---|---|---|
| | Planes | FWHM | $D_{hkl}$ | $D_{avg}$ | D | ε |
| No Capping | (111) | 0.559 | 14.64 | 14.31 | 16.83 | 1.02 |
| | (220) | 0.608 | 14.25 | | | |
| | (311) | 0.641 | 14.04 | | | |
| 0.05 mM PVP | (111) | 0.717 | 11.43 | 11.17 | 11.55 | 0.27 |
| | (220) | 0.806 | 10.74 | | | |
| | (311) | 0.791 | 11.36 | | | |
| 0.10 mM PVP | (111) | 1.128 | 7.27 | 6.57 | 7.53 | 4.31 |
| | (220) | 1.350 | 6.42 | | | |
| | (311) | 1.494 | 6.02 | | | |
| 0.50 mM PVP | (111) | 1.372 | 5.97 | 5.41 | 7.18 | 4.31 |
| | (220) | 1.712 | 5.06 | | | |
| | (311) | 1.732 | 5.20 | | | |

These results align with prior work by Alwany et al. [22], who observed that annealing ZnS nanoparticles up to 340 °C reduced dislocation density and internal strain, promoting crystallinity and lattice relaxation. The hydrothermal conditions employed here achieved similar effects in situ. Unlike the rapid nucleation and growth in co-precipitation, the hydrothermal process allows sufficient time and energy for defect minimization and structural rearrangement, resulting in more ordered nanocrystals. Therefore, showed less peak broadening and lattice parameter shifts under kinetic control, whereas thermally annealed ZnS shows almost stress-free cubic lattices (bulk a ≈5.41 Å) and minimal strain [22].

In Summary, co-precipitation at ambient temperature leads to rapid nucleation and kinetically controlled growth, which traps defects and strain in the lattice. Accordingly, co-precipitated ZnS shows broader XRD peaks and lattice parameters slightly below the bulk value. In contrast, hydrothermal synthesis (high temperature and long time) permits defects to anneal out, yielding more relaxed crystals. Indeed, the hydrothermal ZnS patterns have peak widths near the instrument limit and lattice constants almost identical to bulk ZnS. Moreover, the choice of capping ligand also has a pronounced effect. Large PVP molecules bind more strongly to ZnS surfaces than PEG chains, so they inhibit growth to a greater extent. For example, a 0.50mM PVP-capped sample shows a smaller crystallite size (2.03 nm) as compared to PEG (3.33 nm) at the same capping concentration. This agrees with previous reports (e.g., [39,40]) that PVP stabilizes finer colloids and induces stronger lattice distortion than PEG. In practical terms, PVP's longer chains and stronger surface affinity confer superior growth suppression and yield more monodisperse, high-quality ZnS nanocrystals than PEG.

### 3.2. Morphology and elemental composition

Morphological and elemental analysis were done using the scanning electron microscope (SEM) and energy-dispersive X-ray spectroscopy. Fig. 5 shows the EDX spectra of the synthesised ZnS particles. The presence of Zn and S peaks confirms the formation of ZnS. No peaks related to precursor compounds or other foreign elements were observed in any sample, indicating complete reaction and purity of the final product. The absence of any residual precursor elements further supports the effective synthesis processes.

There is an increase in the concentration of carbon as we increase the polymer concentration in the co-precipitation samples. The presence of carbon is due to the organic polymer, which consists of carbon chains. There is also an increase in the N in the samples synthesised with the PVP due to the presence of the pyrrolidone ring containing nitrogen, suggesting the successful capping of the particles.

The SEM shows the reduced agglomeration in the samples with high polymer capping. Fig. 6 (a) shows ZnS nanoparticles, synthesised without capping agents, gathered into larger clumps or aggregates. The particles appear to be agglomerated together in uneven lumps, indicating a higher degree of agglomeration. In contrast, in Fig. 6 (b) to (e), when the capping was introduced, a much more uniform distribution of the ZnS nanoparticles was observed. Here, the particles are more discrete and spread out, forming smaller clusters with blurry edges. This suggests that the sample with the higher capping concentration has significantly less agglomeration. This agglomeration is reduced more in the PVP-capped sample as compared to the PEG at the same concentration. This difference is due to the longer vinyl chain in the PVP than the PEG, which introduces stronger steric hindrance between the particles. Moreover, the hydrothermal synthesis also shows larger chunks of particles, showing the aggregation due to the high temperature synthesis.

Comparing PEG and PVP capping, PVP proves to be the more effective capping agent for controlling ZnS growth and dispersion. The PVP-capped ZnS samples are markedly smaller in mean size and less aggregated than their PEG-capped counterparts at equivalent capping concentrations. PVP's superior performance can be understood by its strong binding affinity and steric stabilization as described earlier. PVP molecules adsorb to ZnS surfaces via their lactam (C–N and C=O) groups, effectively passivating the surface and impeding further growth or clustering [43]. PEG, while still providing some stabilization, has weaker interaction and shorter chains, so PEG-capped particles show a somewhat larger size and occasional agglomerates relative to the tightly PVP-coated ones. These observations are in line with previous reports that polymers like PVP with long-chain, strongly-coordinating functional groups can yield finer and more monodisperse colloids than less strongly binding capping agents [44,45]. In our samples, the 0.50 mM PVP-capped ZnS exhibits the most uniform dispersion of nanoparticles, with only minimal clustering, whereas the 0.50 mM PEG-capped ZnS, despite being much improved over uncapped ZnS, still contains a few larger clumps. Thus, a high concentration of PVP can effectively curb nanoparticle aggregation and growth, resulting in smaller, more uniformly distributed ZnS QDs, PEG requires higher amounts to achieve a similar level of size control. These SEM results, together with the EDX findings, consistently demonstrate that increasing the capping agent (especially PVP) leads to better surface coverage and particle isolation, which translates to reduced agglomeration and finer particle morphology in the ZnS QDs.

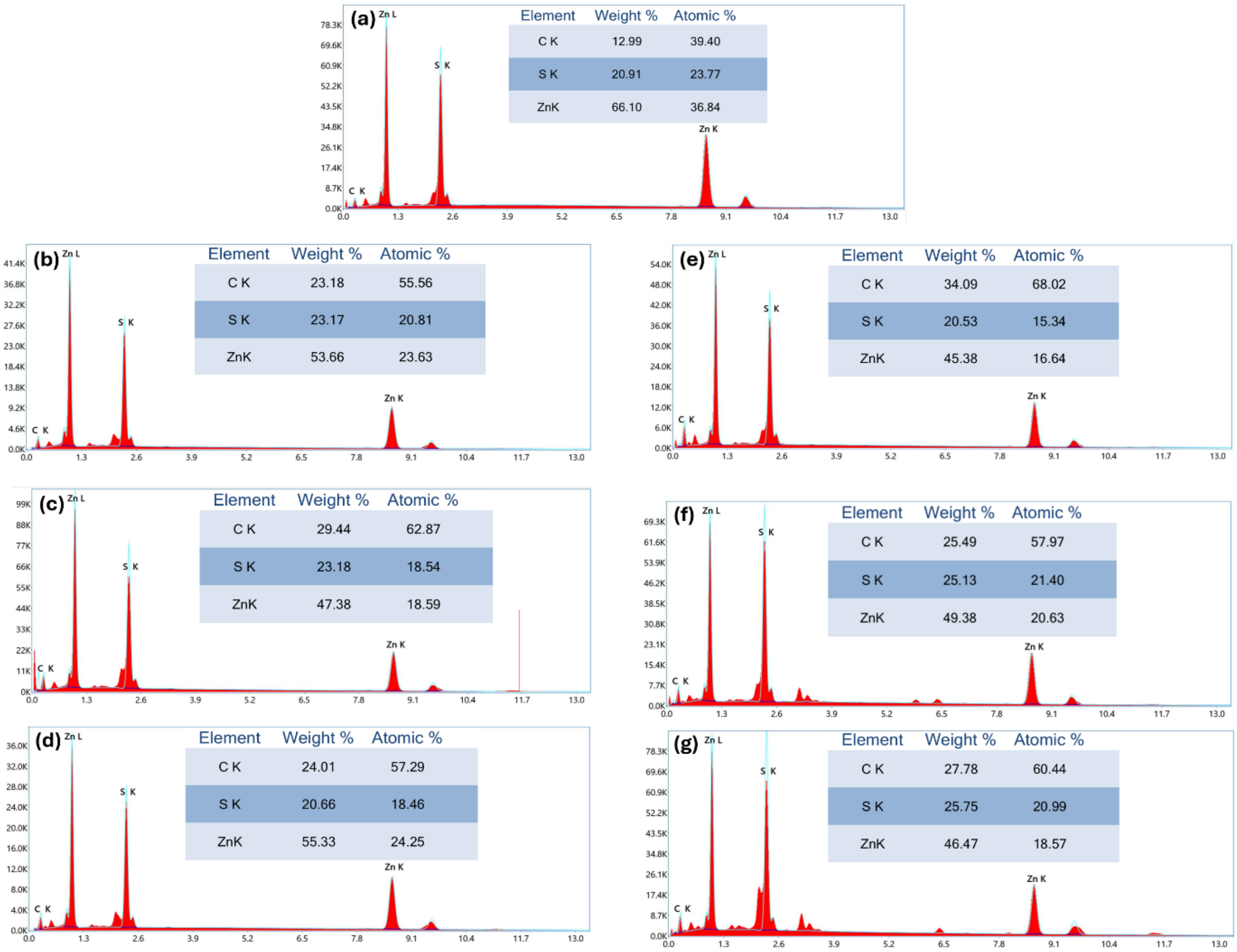

*Fig. 5. EDX spectra of ZnS QDs synthesised by co-precipitation with (a) no capping, (b) 0.05 mM PEG, (c) 0.50 mM PEG, (d) 0.05 mM PVP, (e) 0.50 mM PVP, and hydrothermally synthesised samples with (f) 0.05 mM PVP and (g) 0.50 mM PVP, confirming successful synthesis and capping.*

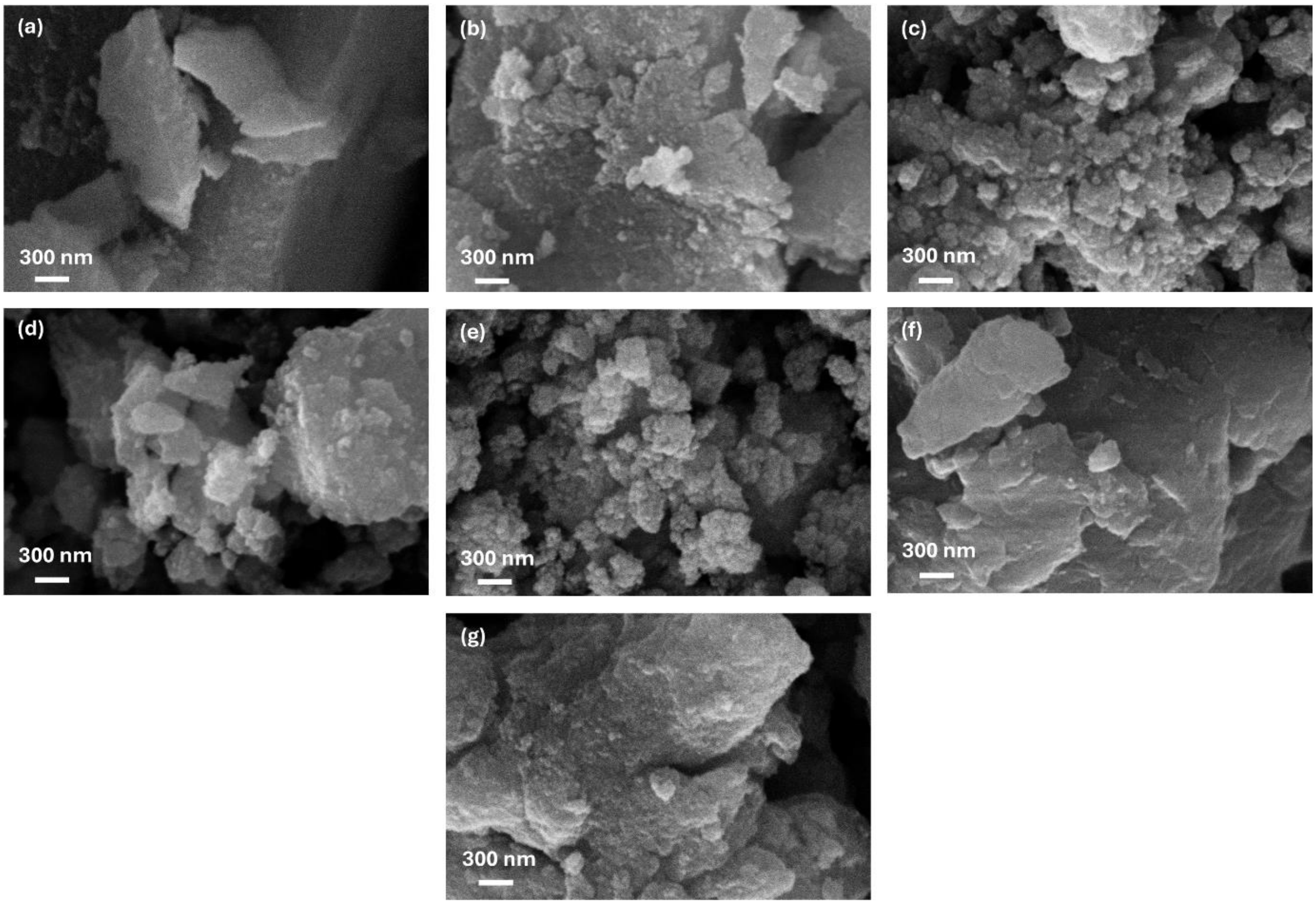

*Fig. 6. SEM images of ZnS nanoparticles synthesised by the co-precipitation method with (a) no capping, (b) 0.05 mM PEG, (c) 0.50 mM PEG, (d) 0.05 mM PVP, (e) 0.50 mM PVP, and hydrothermally synthesised samples with (f) 0.05 mM PVP and (g) 0.50 mM PVP.*

### 3.3. Optical properties

To examine the absorption spectra and the effect of particle size on the bandgap of ZnS QDs, UV-Vis spectroscopy was used. UV-Vis absorption spectra of synthesised ZnS QDs are shown in **Fig. 7**. All samples exhibit a clear absorption edge in the near-UV. It is seen that increasing the polymer content causes the absorption onset to shift to shorter wavelengths (a blue shift). For example, the 0.50 mM PVP sample (green curve) has the most blue-shifted edge, while the sample without capping (black curve) shows a more red-shifted, gradual rise. These shifts mean that high-capped samples have larger optical band gaps. In colloidal QDs, a shorter-wavelength (higher-energy) absorption edge corresponds to a larger band gap caused by quantum confinement. It is also well established that the bandgap increases with decreasing particle size for quantum dots. Thus, the systematic blue shift with higher polymer concentration yields smaller ZnS QDs. This makes sense because the polymer acts as a capping/stabilizing agent and binds itself to surface ions and slows the growth, yielding smaller, more monodisperse particles. The high-capped spectra also display notably sharper absorption edges, i.e., the green and blue curves rise more steeply than the black curve in the co-precipitation samples. An abrupt absorption edge indicates more uniform nanoparticle size distribution, whereas a smeared-out edge (as in the no-capping sample) signals a wider spread of sizes. Literature also reports that increasing PVP concentration produces monodisperse colloidal solutions with sufficiently small size and narrow size distributions [41]. These shifts were not as noticeable in the hydrothermally synthesised samples, as the smallest crystallite size we can achieve by this method was above 5 nm. This is almost twice the Bohr exciton radius of ZnS, so these samples do not show a significant quantum confinement effect.

To quantify the band gap, the absorption data were analysed using Tauc plots of $(\alpha h\nu)^2$ versus $h\nu$. Extrapolating the linear portion of each Tauc curve to the energy axis yields the optical band gap ($E_g$). The Tauc analysis shows a clear trend that $E_g$ increases with the concentration of the capping agents. For example, ZnS without any capping showed $E_g$ = 3.60 eV, and this gap has been raised to $E_g$ = 3.80 eV for the 0.50 mM PVP capped sample, as shown in Table 5.

*Table 5. Optical band gaps estimated from direct-transition Tauc plots for ZnS samples prepared by co-precipitation and hydrothermal synthesis.*

| | Sample | Bandgap |
|---|---|---|
| Co- precipitation Method | (a) No Capping | 3.62 eV |
| | (b) 0.05 mM PEG | 3.68 eV |
| | (c) 0.10 mM PEG | 3.71 eV |
| | (d) 0.50 mM PEG | 3.75 eV |

| | Sample | Bandgap |
|---|---|---|
| | (e) 0.05 mM PVP | 3.68 eV |
| | (f) 0.10 mM PVP | 3.73 eV |
| | (g) 0.50 mM PVP | 3.81 eV |
| Hydrothermal Method | (h) 0.05 mM PVP | 3.61 eV |
| | (i) 0.10 mM PVP | 3.61 eV |
| | (j) 0.50 mM PVP | 3.61 eV |

Overall, all synthesised ZnS QDs exhibit blue-shifted absorption edges relative to bulk ZnS ($E_g$ bulk ~3.60 eV)[46]. The Tauc-derived band gap rises as crystallite size falls, in line with quantum confinement. For instance, the smallest PVP-capped QDs (D~2.03 nm) have Eg ~ 3.80 eV, a +0.20 eV shift above the bulk bandgap of 3.60 eV. PEG-capped QDs (~4.3 nm) show $E_g$ ~ 3.71 eV. Tauc-plot analysis confirms that adding more polymer as a capping agent yields smaller, more uniformly sized ZnS QDs, leading to stronger quantum confinement and larger optical band gaps. All this entirely agrees with the literature that smaller crystallites tend to have larger band gaps [47]. Conversely, hydrothermally synthesised enlarged crystals do not exhibit a noticeable blue shift in the absorption spectra of ZnS QDs due to their crystallite size above 5.81 nm. Chukwuocha et al. reported that materials like CdSe, ZnS, and GaAs show no confinement effect when the particle's diameter is roughly twice the Bohr exciton radius, and a strong quantum confinement effect is observed when the size is reduced to less than the Bohr exciton radius [4]. These trends are quantitatively consistent with Brus's effective-mass model, which states that $E_g$ increases sharply when D approaches the exciton Bohr diameter [48]. Thus, the systematic Tauc analysis confirms that the choice of synthesis route and ligand dictates the degree of confinement, and kinetically grown, PVP-capped ZnS QDs (small and strained) attain the largest band gaps, while thermally grown QDs (larger and relaxed) display narrower gaps closer to the bulk.

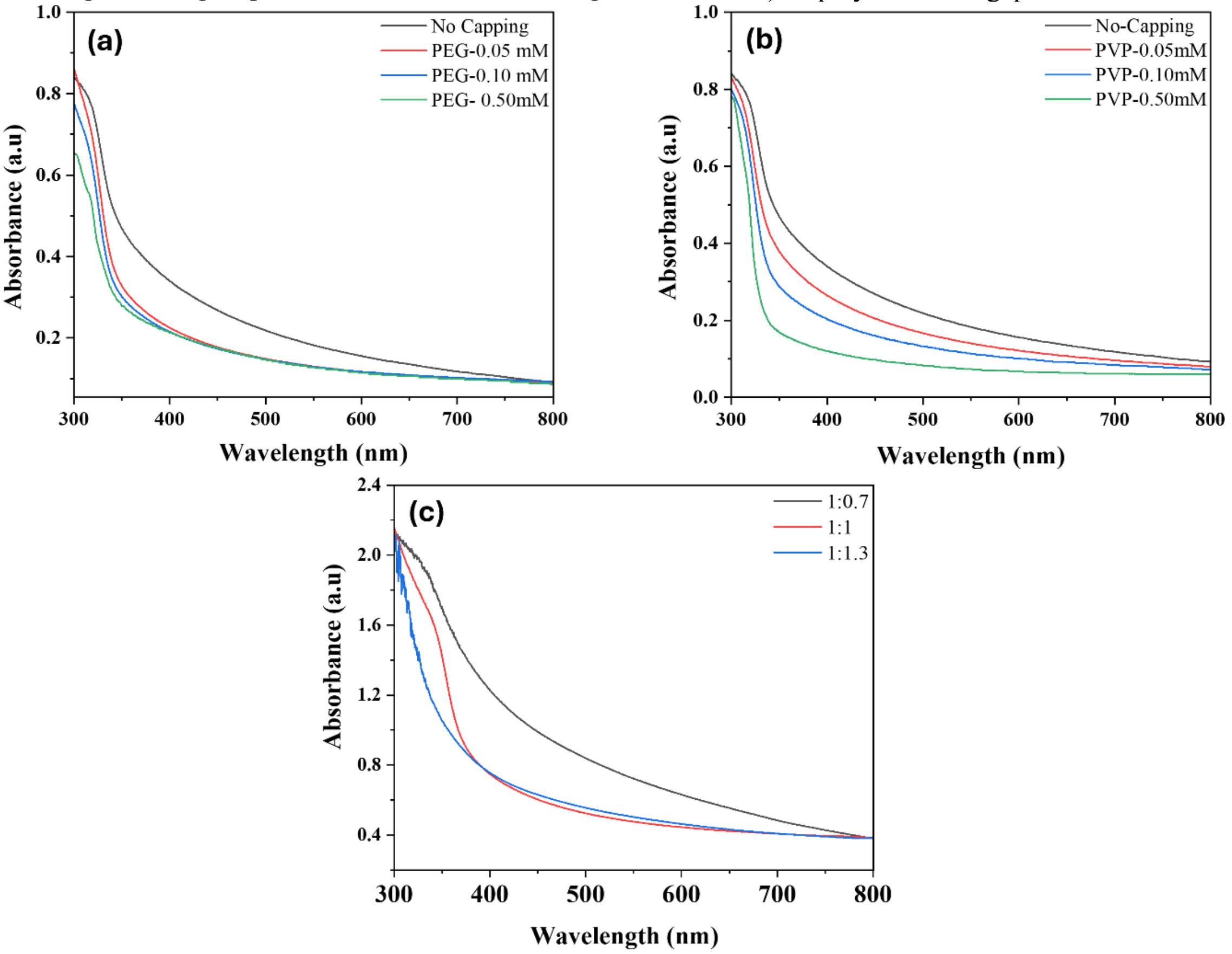

*Fig. 7. UV–Vis absorption spectra of ZnS nanoparticles in the UV range: (a, b) co-precipitated samples showing a strong quantum-confinement effect and a noticeable blue shift, and (c) hydrothermally synthesised nanoparticles showing less blue shift because of their larger particle size.*

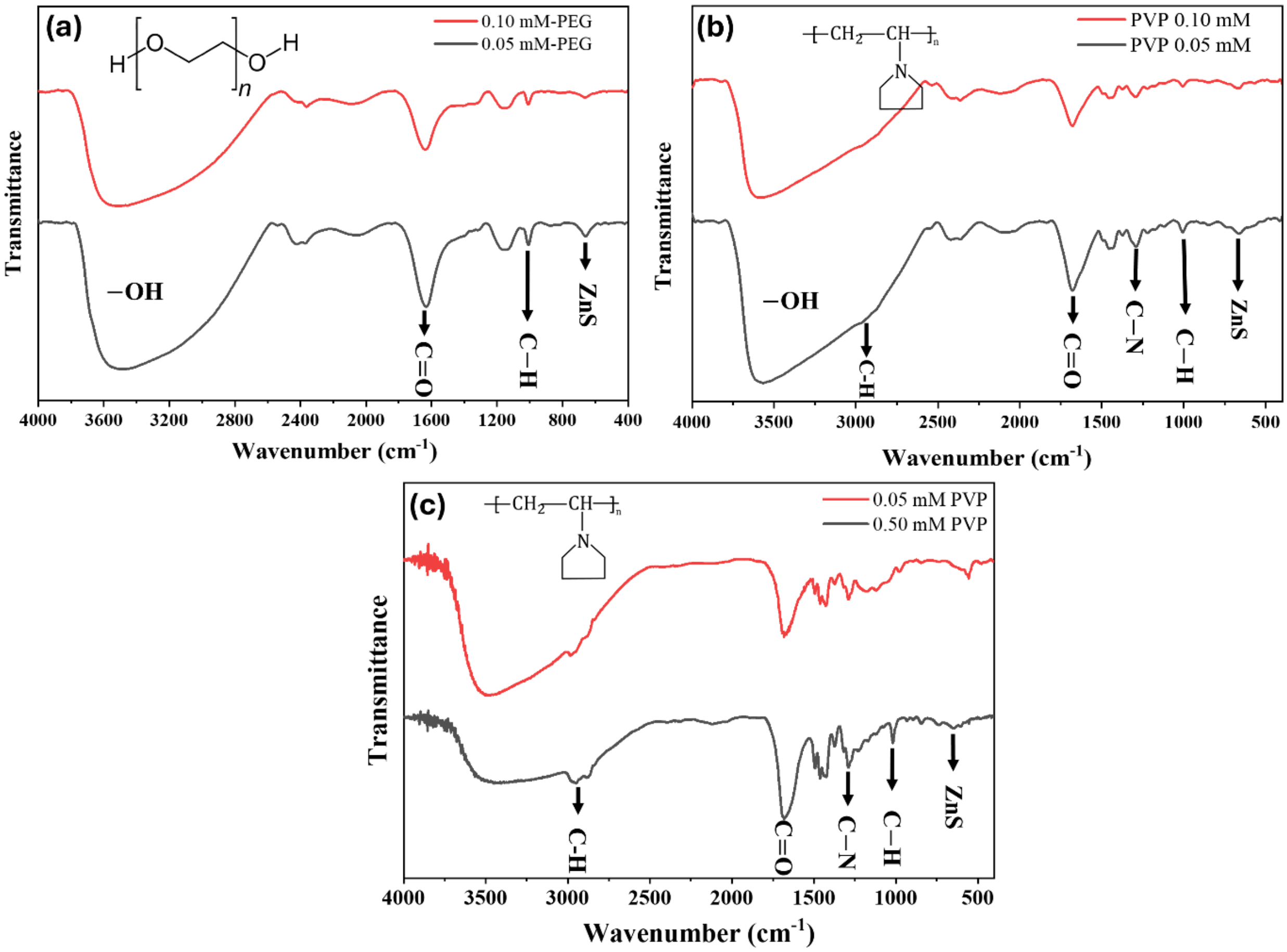

*Fig. 8. Fourier-transform infrared (FTIR) spectra of ZnS QDs showing characteristic vibrational bands of PVP and PEG. Peaks near 610–640 cm$^{-1}$ confirm Zn–S bonding; polymer-specific bands confirm surface functionalisation.*

### 3.4. Surface chemistry analysis

FTIR spectra of ZnS QDs, given in Fig. 8, show the characteristic vibrational bands of PEG and PVP. A broad band around $\approx$3430 cm$^{-1}$ is present (O–H stretch from adsorbed water) [49]. The C–H stretching bands appear near 2960 and 2870 cm$^{-1}$ from the presence of $CH_2$ in both structures [49]. A prominent peak at $\approx$1650-1680 cm$^{-1}$ corresponds to the carbonyl (C=O) stretch of the pyrrolidone ring. We also see C-H bending modes around 1420 cm$^{-1}$ ($CH_2$ scissoring) and a C–N (lactam) stretch near 1280-1300 cm$^{-1}$. These bands confirm the presence of polymer on the sample (their positions match literature values) [49].

In the low-frequency fingerprint region (<1000 cm$^{-1}$), bands attributable to Zn–S bonds were observed. For example, ZnS typically shows Zn–S stretching modes around 410-490 cm$^{-1}$ (symmetric/asymmetric stretches) and 610-640 cm$^{-1}$ [50]. One can identify a band near 610-640 cm$^{-1}$ in this spectrum, consistent with reported ZnS modes [50]. Some reports also cite peaks at 509 and 612 cm$^{-1}$ for cubic ZnS, which align with the features here [51,52]. These low-wavenumber bands confirm the Zn–S framework of the quantum dots.

The key difference between PEG and PVP is the presence of a C–N stretching peak near ~1280 cm$^{-1}$, which is absent in PEG. Similarly, a small peak around ~2960 cm$^{-1}$, representing C–H stretching, is also missing in the PEG samples. This may be due to the stronger presence of –OH bands in PEG, as its chains typically end with –OH groups, leading to more pronounced –OH vibrations compared to PVP.

### 3.5. Hydrodynamic size distribution

Dynamic light scattering was used to examine the hydrodynamic size distributions of the ZnS dispersions, as shown in Fig. 9. The smallest reported hydrodynamic diameter was 15.03 nm for a co-precipitated sample. This size is actually the hydrodynamic size of the particle. It includes the effects of vibrations of the particles as they are in Brownian motion, and a thick layer of polymer surrounding these particles. A full layer thickness of PVP (molar mass = 40000 g/mol) in adsorbed form may range from 5 to 10 nm, depending on stretching, surface interaction, and solvent conditions. Similar has been reported in the literature [53,54]. Therefore, our estimated particle size could be around 5 nm, suggesting the crystallite size calculated from the XRD spectra is actually the size of the individual particle.

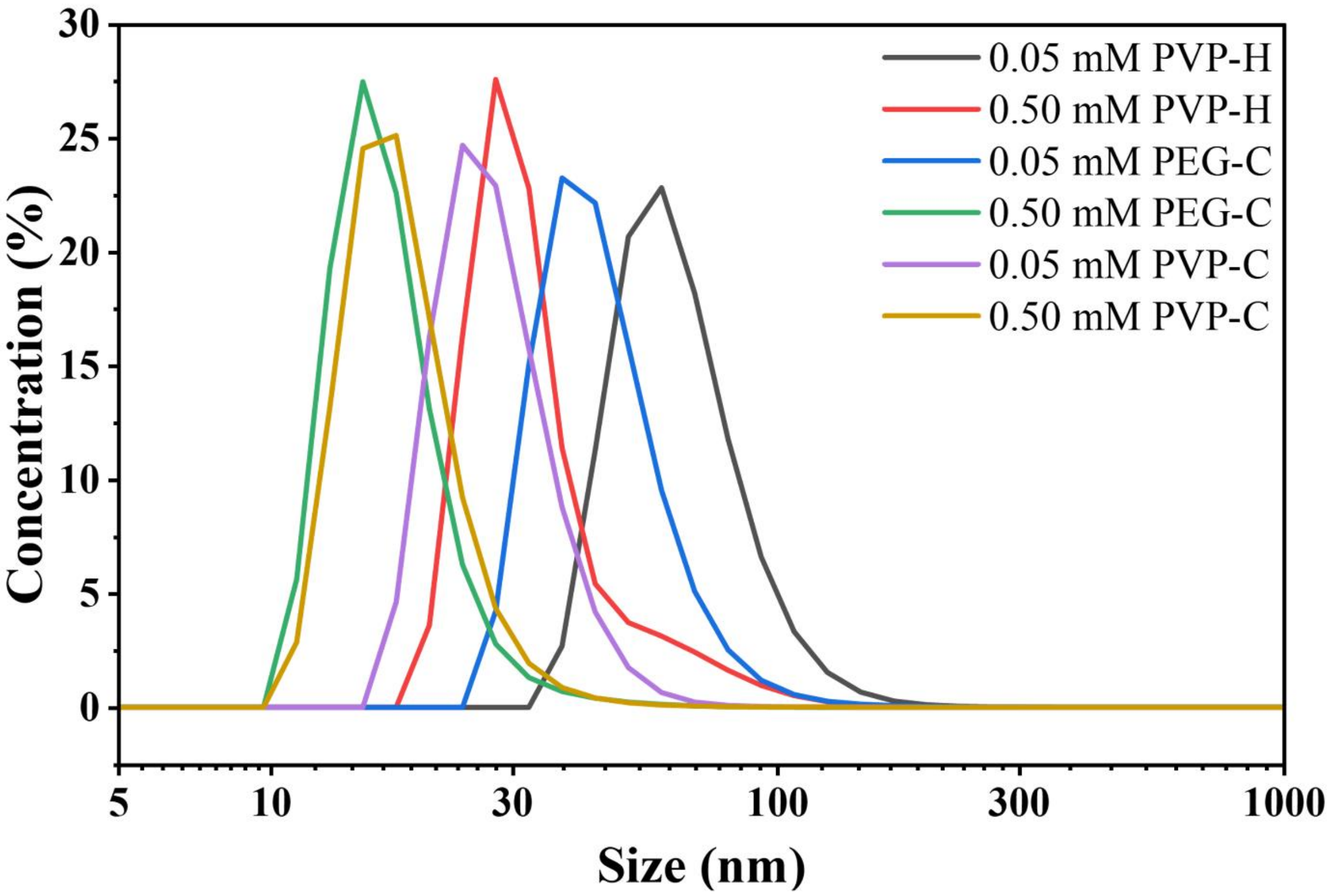

*Fig. 9. Dynamic light scattering (DLS) analysis of ZnS QDs showing particle size distribution by number. The -H in the legend represents the samples synthesised by the hydrothermal method, and -C represents the samples synthesised by the co-precipitation method.*

In summary, the DLS results showed that increasing the capping concentration led to a narrower particle size distribution. Samples with lower capping concentrations exhibited polydispersity in particle size. Furthermore, these findings support our XRD results, suggesting that the calculated crystallite sizes closely represent the actual particle size of the synthesised ZnS QDs [53,54], [55]. Hydrothermal ZnS, though larger, also showed narrower distributions even with low capping, likely due to controlled nucleation and enhanced crystallinity under high-temperature, high-pressure conditions [56]. Thus, both polymer concentration and synthesis route significantly influence particle size and size uniformity.

These results show that the polymer concentration and synthesis route influenced both crystalline-domain growth and dispersion behaviour. However, DLS does not independently establish the diameter of individual ZnS crystallites. In particular, subtracting an assumed polymer-shell thickness from the hydrodynamic diameter would introduce considerable uncertainty because polymer conformation depends on molecular weight, surface coverage, solvent and particle aggregation.

Considering the results together, room-temperature co-precipitation and hydrothermal treatment produced distinctly different structure–property relationships. Rapid precipitation in the presence of polymer restricted the growth of coherent ZnS domains, with the strongest reduction observed for the 0.50 mM PVP condition. These small domains showed broader XRD reflections, larger estimated microstrain and higher optical band gaps. Hydrothermal treatment, by contrast, produced larger and comparatively relaxed crystalline domains with lattice parameters and optical band gaps closer to those of bulk ZnS. Polymer addition still restricted hydrothermal crystal growth, but its effect on the optical band gap was limited. The agreement between the XRD and UV–visible trends supports the conclusion that the synthesis route and polymer conditions provide control over ZnS crystallite growth and optical absorption. At the same time, SEM and DLS show that the final products contain polymer-associated aggregates; therefore, coherent crystallite size, hydrodynamic size and aggregate morphology should be treated as complementary rather than interchangeable measurements.

## 4. Conclusions

We have demonstrated that simple, low-temperature synthesis enables systematic control of ZnS QDs properties. Comparing hydrothermal and ambient co-precipitation methods revealed that aqueous precipitation at room temperature (with PVP/PEG capping) produces uniformly small, crystalline ZnS QDs (~3-5 nm) with tuneable optical band gaps. In particular, polymer capping effectively limited growth to yield monodisperse nanocrystals, whereas hydrothermal processing led to much larger particles. The resulting band gaps (up to ~3.80 eV for the smallest QDs) were consistent with strong quantum confinement. Importantly, these syntheses avoided harsh chemicals and high heat, underscoring the sustainability of the approach. By using benign precursors and ambient conditions, this work illustrates a viable route to synthesising ZnS QDs for practical nanomaterials applications. Such room-temperature, low-toxicity methods could be extended to other nanosystems, advancing the development of green, large-scale nanoparticle manufacturing.

**Acknowledgements.** The authors acknowledge the funding from the GIK Institute.

**Author contributions.** Rao Uzair Ahmad performed experiments, analysed results, and wrote the initial draft of the manuscript. Nasir Javed supervised the project, revised, and corrected the manuscripts.

**Funding.** GIK Institute provided funding for acquiring consumables.

**Data availability.** Experimental data, including results of all characterisation, are available and can be provided on request.

**Conflict of interest.** The authors declare no competing interests.

## References

[1] Q. Cao, J. Feng, K. T. Chang, W. Liang, H. Lu, Adv. Mater., **37**(23), 2409096 (2025).
[2] S. C. Lims, N. A. Tran, V. D. Dao, P. V. Pham, Coord. Chem. Rev., **528**, 216423 (2025).
[3] O. E. Semonin, J. M. Luther, M. C. Beard, Mater. Today, **15**(11), 508 (2012).
[4] E. O. Chukwuocha, M. C. Onyeaju, T. S. T. Harry, World J. Condens. Matter Phys., **2**(2), 96 (2012).
[5] B. Van Thang, H. T. Tung, D. H. Phuc, T. P. Nguyen, T. Van Man, L. Q. Vinh, Sol. Energy Mater. Sol. Cells, **250**, 112042 (2023).
[6] S. K. Mishra, R. K. Srivastava, S. G. Prakash, R. S. Yadav, A. C. Panday, Electron. Mater. Lett., **7**(1), 31 (2011).
[7] C. Tyagi, A. Sharma, R. Kurchania, J. Non-Oxide Glasses, **6**(2), 23 (2014).
[8] D. Nath, F. Singh, R. Das, Mater. Chem. Phys., **239**, 122021 (2020).
[9] P. Kumar, Mater. Today Proc., **51**, 900 (2022).
[10] V. Lahariya, M. Michalska-Domańska, S. J. Dhoble, "Synthesis, structural properties, and applications of cadmium sulphide quantum dots", in "Quantum Dots", Elsevier, Amsterdam, p. 235 (2023).
[11] B. P. Jackson, D. Bugge, J. F. Ranville, C. Y. Chen, Environ. Sci. Technol., **46**(10), 5550 (2012).
[12] H. M. Mistry, M. P. Deshpande, A. B. Hirpara, N. M. Suchak, S. H. Chaki, S. V. Bhatt, Opt. Mater., **159**, 116529 (2025).
[13] N. Kaur, S. Kaur, J. Singh, M. Rawat, J. Bioelectron. Nanotechnol., **1**(1), 1 (2016).
[14] S. Saleem, S. Khalid, M. A. Malik, A. Nazir, Energy Fuels, **38**(11), 9153 (2024).
[15] J. C. Melendres-Sánchez, R. López-Delgado, G. Saavedra-Rodríguez, R. C. Carrillo-Torres, R. Sánchez-Zeferino, A. Ayón, M. E. Álvarez-Ramos, J. Mater. Sci.: Mater. Electron., **32**(2), 1457 (2021).
[16] M. Kim, D. Kim, O. Kwon, H. Lee, Micromachines, **13**(2), 269 (2022).
[17] N. Verma, A. K. Singh, N. Saini, Sens. Bio-Sens. Res., **15**, 41 (2017).
[18] C. Y. Huang, D. Y. Wang, C. H. Wang, Y. T. Chen, Y. T. Wang, Y. T. Jiang, Y. J. Yang, C. C. Chen, Y. F. Chen, ACS Nano, **4**(10), 5849 (2010).
[19] B. Li, A. H. Amin, A. M. Ali, M. Isam, A. A. Lagum, M. M. Sabugaa, R. D. C. Pecho, H. M. Salman, M. F. Nassar, Chemosphere, **336**, 139208 (2023).
[20] H. Labiadh, K. Lahbib, S. Hidouri, S. Touil, T. B. Chaabane, Asian Pac. J. Trop. Med., **9**(8), 757 (2016).
[21] T. Mangeolle, I. Yakavets, S. Marchal, M. Debayle, T. Pons, L. Bezdetnaya, F. Marchal, Nanomaterials, **8**(8), 572 (2018).
[22] A. B. Alwany, G. M. Youssef, O. M. Samir, M. A. Algradee, N. A. A.Yahya, M. A. Swillam, S. Humaidi, R. Abd-Shukor, Sci. Rep., **13**(1), 10314 (2023).
[23] L. Dai, R. Lesyuk, A. Karpulevich, A. Torche, G. Bester, C. Klinke, J. Phys. Chem. Lett., **10**(14), 3828 (2019).
[24] A. Bala, R. Sehrawat, A. K. Sharma, P. Soni, J. Mater. Sci.: Mater. Electron., **32**(12), 16382 (2021).
[25] B. Sarangi, S. P. Mishra, N. Behera, Mater. Sci. Semicond. Process., **147**, 106723 (2022).
[26] A. M. Salaheldin, J. Walter, P. Herre, I. Levchuk, Y. Jabbari, J. M. Kolle, C. J. Brabec, W. Peukert, D. Segets, Chem. Eng. J., **320**, 232 (2017).
[27] Y. Hu, B. Hu, B. Wu, Z. Wei, J. Li, J. Mater. Sci.: Mater. Electron., **29**(19), 16715 (2018).
[28] H. Peng, B. Liuyang, Y. Lingjie, L. Jinlin, Y. Fangli, C. Yunfa, Nanoscale Res. Lett., **4**(9), 1047 (2009).
[29] E. M. Sala, M. Godsland, Y. I. Na, A. Trapalis, J. Heffernan, Nanotechnology, **33**(6), 065601 (2022).
[30] D. V. Talapin, A. L. Rogach, A. Kornowski, M. Haase, H. Weller, Nano Lett., **1**(4), 207 (2001).
[31] X. Peng, L. Manna, W. Yang, J. Wickham, E. Scher, A. Kadavanich, A. P. Alivisatos, Nature, **404**(6773), 59 (2000).
[32] V. Sabaghi, F. Davar, Z. Fereshteh, Ceram. Int., **44**(7), 7545 (2018).
[33] K. Agarwal, H. Rai, S. Mondal, Mater. Res. Express, **10**(6), 062001 (2023).
[34] P. Iranmanesh, S. Saeednia, M. Nourzpoor, Chin. Phys. B, **24**(4), 046104 (2015).
[35] M. A. Behrens, A. Franzén, S. Carlert, U. Skantze, L. Lindfors, U. Olsson, Soft Matter, **21**(12), 2349 (2025).
[36] R. Javed, A. Sajjad, S. Naz, H. Sajjad, Q. Ao, Int. J. Mol. Sci., **23**(18), 10521 (2022).
[37] A. Kalita, M. P. C. Kalita, Physica E, **92**, 36 (2017).
[38] M. Goswami, Mater. Lett., **372**, 137029 (2024).
[39] B. Ajitha, Y. A. Kumar Reddy, P. S. Reddy, H. J. Jeon, C. W. Ahn, RSC Adv., **6**(42), 36171 (2016).
[40] O. Wiranwetchayan, S. Promnopat, T. Thongtem, A. Chaipanich, S. Thongtem, Mater. Chem. Phys., **240**, 122219 (2020).
[41] N. Soltani, E. Saion, M. Erfani, K. Rezaee, G. Bahmanrokh, G. P. C. Drummen, A. Bahrami, M. Z. Hussein, Int. J. Mol. Sci., **13**(10), 12412 (2012).
[42] K. Chari, W. C. Lenhart, J. Colloid Interface Sci., **137**(1), 204 (1990).
[43] I. A. Safo, M. Werheid, C. Dosche, M. Oezaslan, Nanoscale Adv., **1**(8), 3095 (2019).

[44] R. Sharma, M. J. R., S. Thakur, S. Govindasamy, R. B. Singh, H. C. Swart, J. Prakash, J. Mater. Chem. C, **13**(25), 12870 (2025).
[45] G. Vinoth, B. Janarthanan, J. Sahadevan, A. Dinesh, M. Santhamoorthy, S. Santhoshkumar, Mater. Sci. Eng. B, **322**, 118602 (2025).
[46] P. D'Amico, A. Calzolari, A. Ruini, A. Catellani, Sci. Rep., **7**(1), 16805 (2017).
[47] P. V. Joglekar, D. J. Mandalkar, M. A. Nikam, N. S. Pande, A. Dubal, Int. J. Res. Advent Technol., **7**(1), 510 (2019).
[48] L. E. Brus, J. Chem. Phys., **80**(9), 4403 (1984).
[49] A. M. Abdelghany, M. S. Mekhail, E. M. Abdelrazek, M. M. Aboud, J. Alloys Compd., **646**, 326 (2015).
[50] S. Sheshmani, M. Mardali, J. Polym. Environ., **32**(11), 5783 (2024).
[51] P. Kumari, K. P. Misra, S. Samanta, A. Rao, A. Bandyopadhyay, S. Chattopadhyay, J. Lumin., **251**, 119258 (2022).
[52] M. Jothibas, C. Manoharan, S. Johnson Jeyakumar, P. Praveen, I. Kartharinal Punithavathy, J. Prince Richard, Sol. Energy, **159**, 434 (2018).
[53] A. A. Salam, S. Ebrahim, M. Soliman, A. Shokry, Sci. Rep., **14**(1), 20986 (2024).
[54] F. N. S. Neto, L. A. Morais, L. F. Gorup, L. S. Ribeiro, T. J. Martins, T. Y. Hosida, P. Francatto, D. B. Barbosa, E. R. Camargo, A. C. B. Delbem, Colloids Interfaces, **7**(4), 66 (2023).
[55] M. Tejamaya, I. Römer, R. C. Merrifield, J. R. Lead, Environ. Sci. Technol., **46**(13), 7011 (2012).
[56] C. Martins, C. Rolo, V. R. G. Cacho, L. C. J. Pereira, J. P. Borges, J. C. Silva, T. Vieira, P. I. P. Soares, Mater. Adv., **6**(5), 1726 (2025).

**Graphical Abstract**

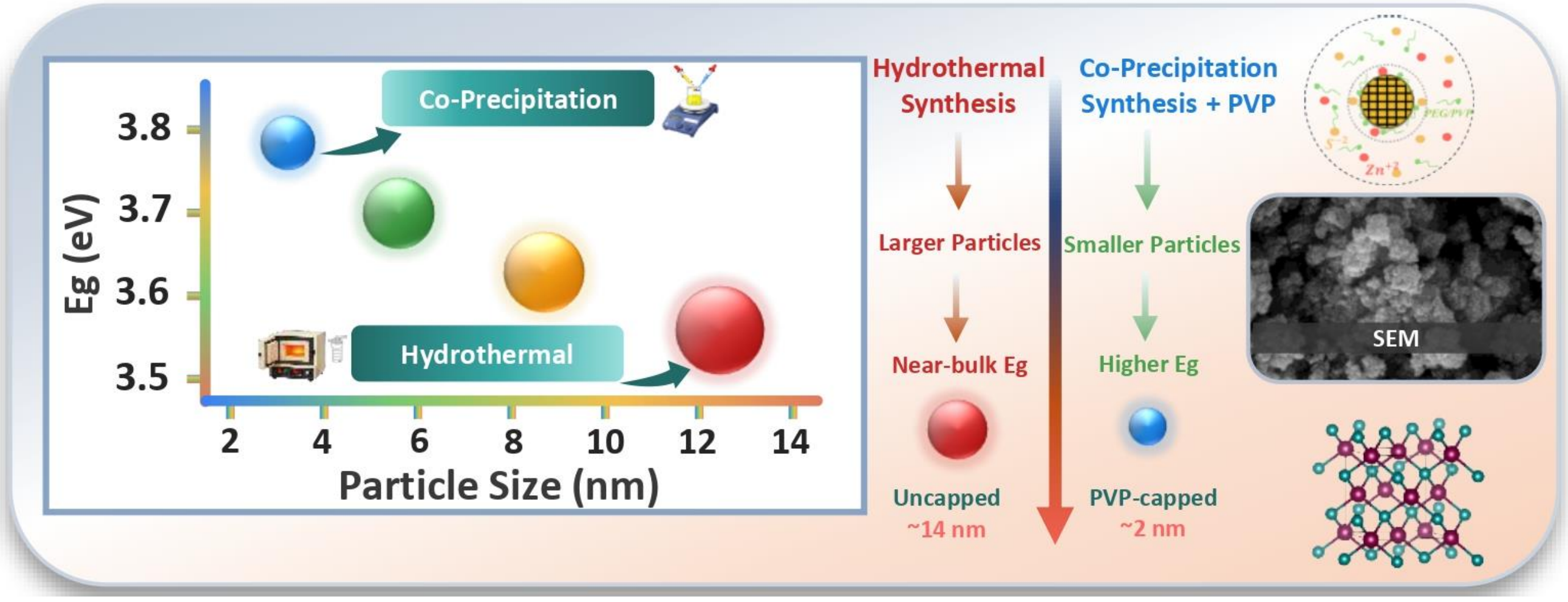